\documentclass[prb,preprint]{revtex4-1} 

\usepackage{amsmath} 

\usepackage{graphicx} 

\begin{document}

\title{A Jumping Cylinder on an Incline}
\author{R.W. G\'omez}
\email{rgomez@unam.mx} 
\author{J.J. Hern\'andez-G\'omez}
\email{jorge\_hdz@ciencias.unam.mx}
\author{V. Marquina}
\email{marquina@unam.mx} 
\affiliation{Facultad de Ciencias, Universidad Nacional Aut\'onoma de M\'exico\\
M\'exico, D.F. 04510}
\date{\today}

\begin{abstract}
The problem of a cylinder of mass $m$ and radius $r$, with its center of mass out of the cylinder axis, rolling on an incline that makes an angle $\alpha$ with respect to the horizontal is analyzed. The equation of motion is partially solved to obtain the site where the cylinder loses contact with the incline (jumps). Several simplifications are made: the analyzed system consists of an homogeneous disc with a one dimensional straight line of mass parallel to the disc axis at a distance $y < r$ of the center of the cylinder. To compare our results with experimental data, we use a Styrofoam cylinder to which a long brass rod was imbibed parallel to the disc axis at a distance $y < r$ from it, so the center of mass lies at a distance $d$ from the center of the cylinder. Then the disc rolls without slipping on a long wooden ramp inclined at $15^\circ$, $30^\circ$ and $45^\circ$ respect to the horizontal. To determine the jumping site, the motion was recorded with a high-speed video camera (Casio EX~ZR100) at 200 and 480 frames per second. The experimental results agree well with the theoretical predictions. 
\end{abstract}

\maketitle

\section{Introduction}
The motion of an homogeneous cylinder rolling without slipping on an horizontal surface, or on an incline, is a classical problem in most textbooks of Mechanics, but only a few of them address the same problems when the center of mass of the cylinder is out of its axis; actually, we only know two examples were this case is put forward as an end-of-the-chapter exercise.$^{\cite{ref1, ref2}}$ Moreover, we have only found three papers that deal with similar movements for symmetrical inhomogeneous bodies.$^{\cite{ref3, ref4, ref5}}$ To our knowledge, this is the first time that the stated problem is theoretical solved and experimentally corroborated.

\section{Theoretical solution}

\begin{figure}[b!]
\centering
\includegraphics[width=12.0 cm]{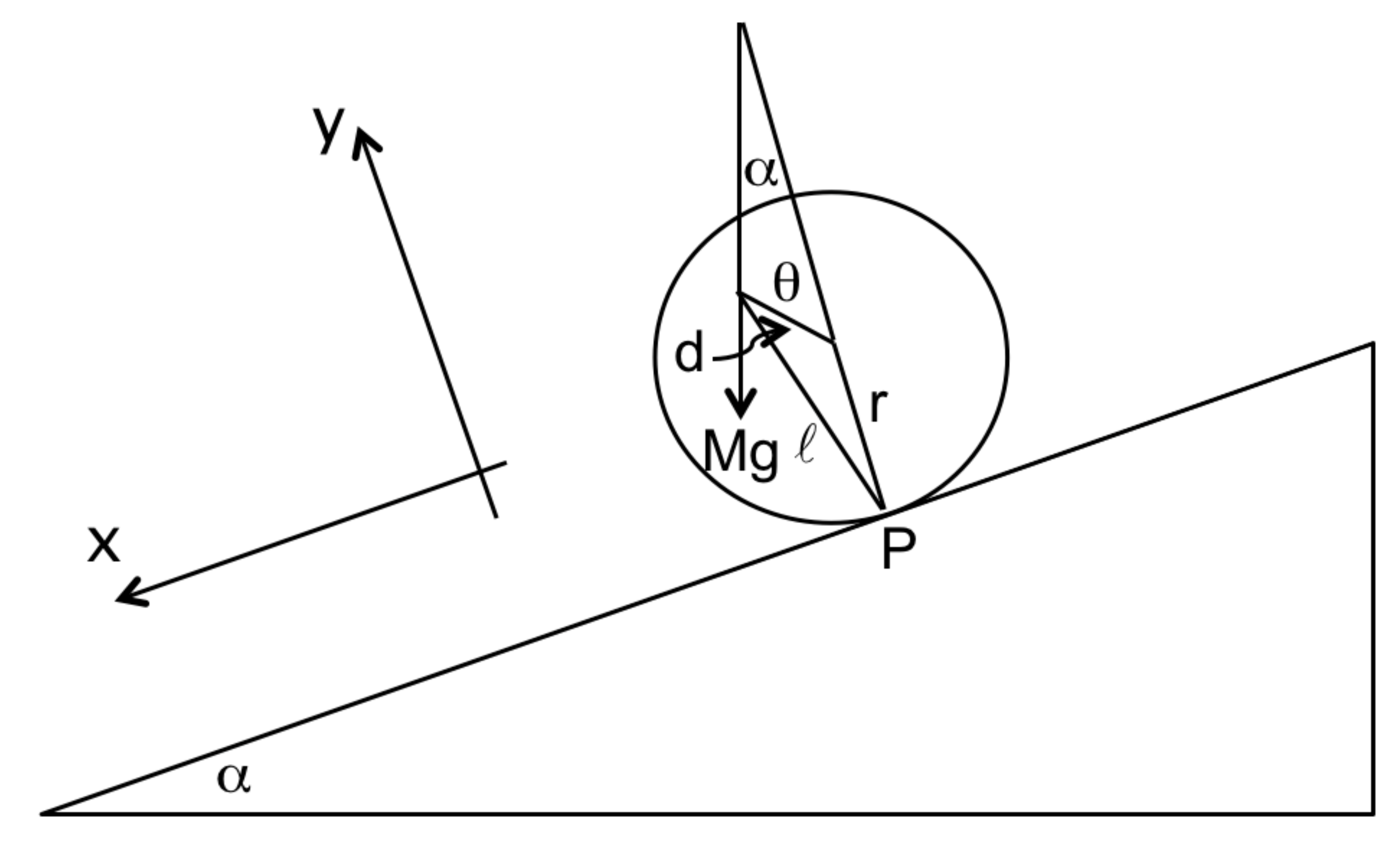}
\caption{The rolling cylinder at an arbitrary position}
\label{f1}
\end{figure}

The simplest way to partially solve the equation of motion is through energy conservation. After the cylinder has roll a distance $x = r\theta$ along the incline plane, its center has descended a distance $h_1 = r\theta\sin\alpha$ and the center of mass has descended a distance $h_2 = d - d\cos\theta\cos\alpha$, so the change $\Delta U$ in the potential energy is:

\begin{equation}
\Delta U =  - Mg[r\theta\sin\alpha+d(1-\cos\theta)\cos\alpha],  \label{E1}
\end{equation}
where $M = m_1 + m_2$ is the total mass of the cylinder and $g = 9.78$ m/s$^2$ is the value of the acceleration of gravity at the place where the experiments are done. 

If the initial conditions are such that the cylinder is at rest and the center of mass lies on a line perpendicular to the incline plane, then the change $\Delta K$ in the kinetic energy of the disc, after it has rotated an angle $\theta$, is:

\begin{equation}
\Delta K = \frac{1}{2}I_P\dot{\theta}^2, \label{E2}
\end{equation}
where $I_P$ is the cylinderÕs moment of inertia respect to the instantaneous axis of rotation $P$. 

Taking a reference frame in which the $x$-axis is along the incline plane, the $y$-axis perpendicular to the same plane and the $z$-axis perpendicular to the $x-y$ plane, the following relation holds between the different quantities involved in the problem (fig. \ref{f1}): 

\begin{equation}
\boldsymbol{\ell}=\mathbf{r}+\mathbf{d}= \boldsymbol{\imath}d\sin\theta + \boldsymbol{\jmath}(r+d\cos\theta), \label{E3}
\end{equation}
where $\boldsymbol{\ell}$ is the vector from the center of mass to the instantaneous axis of rotation $P$, and $\theta$ the rotation angle. The magnitude of $\boldsymbol{\ell}$ is:

\begin{equation}
\ell=\sqrt{r^2+d^2+2rd\cos\theta}, \label{E4}
\end{equation}
so then $I_P$, in virtue of the parallel axes theorem, is:

\begin{equation}
I_P = I_{CM}+M\ell^2=\frac{1}{2}M_1r^2+M(r^2+d^2+2rd\cos\theta).  \label{E5}
\end{equation}

Combining Eq. \eqref{E1} and Eq. \eqref{E2} one obtains:

\begin{equation}
\dot{\theta}^2=2Mg\;\frac{d(1-\cos\theta)\cos\alpha+r\theta\sin\alpha}{\frac{1}{2}M_1r^2+M(r^2+d^2+2rd\cos\theta)}.  \label{E6}
\end{equation}

The condition the cylinder must satisfy in order to lose contact with the incline is that the normal to the incline component of the centrifugal force,

\begin{equation}
\mathbf{F}_C=-M\boldsymbol{\omega}\times(\boldsymbol{\omega}\times\mathbf{d})= M (\boldsymbol{\imath}d\sin\theta + \boldsymbol{\jmath}d\cos\theta)\dot{\theta}^2,  \label{E7}
\end{equation}
equals the normal component of the cylinder total weight, that is:

\begin{equation}
Md\cos\theta\left\{ \frac{d(1-\cos\theta)\cos\alpha+r\theta\sin\alpha}{\frac{1}{2}M_1r^2+M(r^2+d^2+2rd\cos\theta)} \right\} = Mg\cos\alpha,  \label{E8}
\end{equation}
from which the following expression can be obtained:

\begin{equation}
2M\left\{ \frac{d(1-\cos\theta)\cos\alpha+r\theta\sin\alpha}{\frac{1}{2}M_1r^2+M(r^2+d^2+2rd\cos\theta)} \right\} = \frac{\cos\alpha}{d\cos\theta}.  \label{E9}
\end{equation}

\begin{figure}[h!]
\centering
\includegraphics[width=14.5 cm]{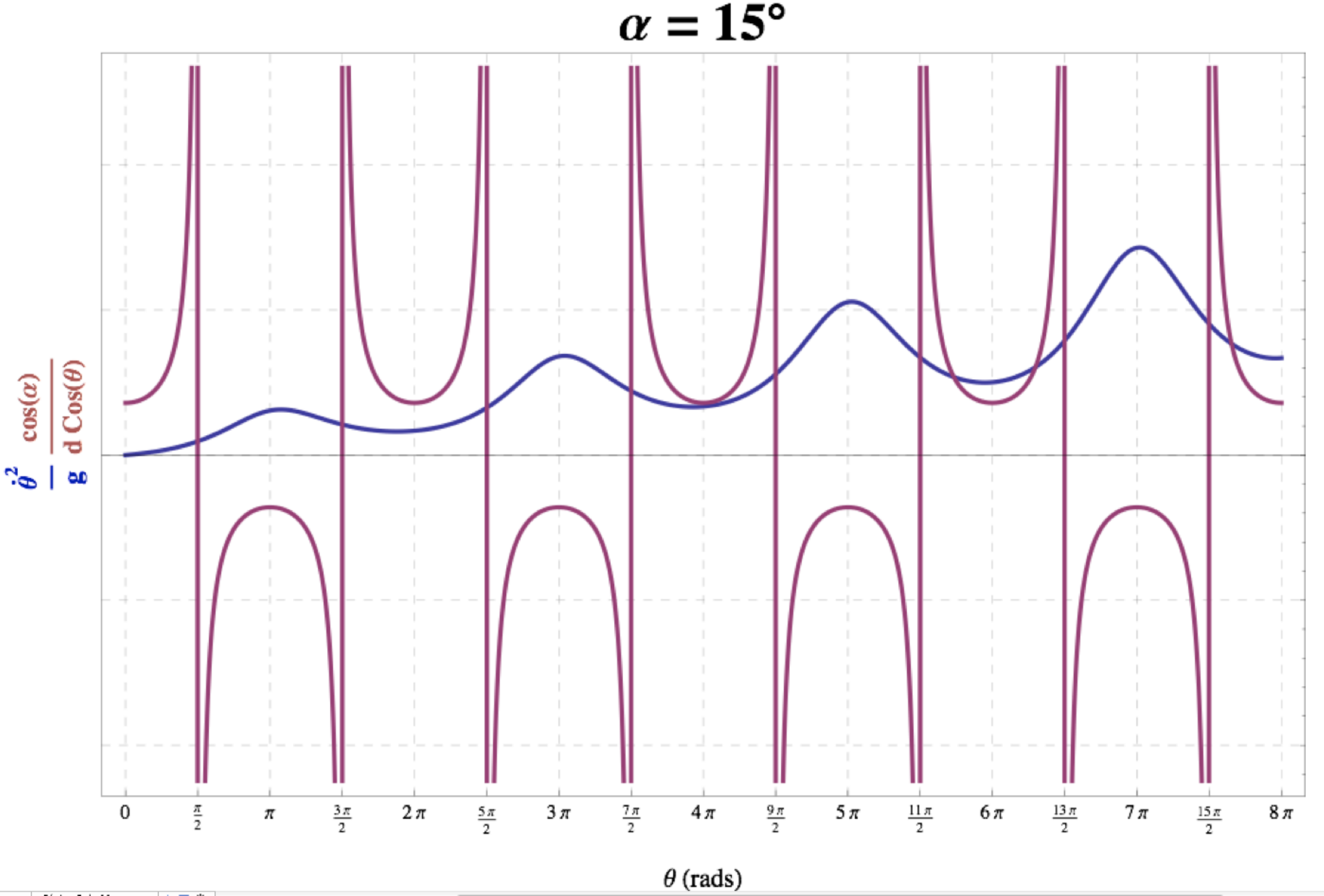}
\caption{Plot of $\dot{\theta}^2/g$ vs $\cos\alpha/d\cos\theta$, for $\alpha=15^\circ$. The first intersection is at $\theta = 18.01$ rads, so the cylinder jumps at $180.1$ cm.}
\label{f2}
\end{figure}

\begin{figure}[h!]
\centering
\includegraphics[width=13 cm]{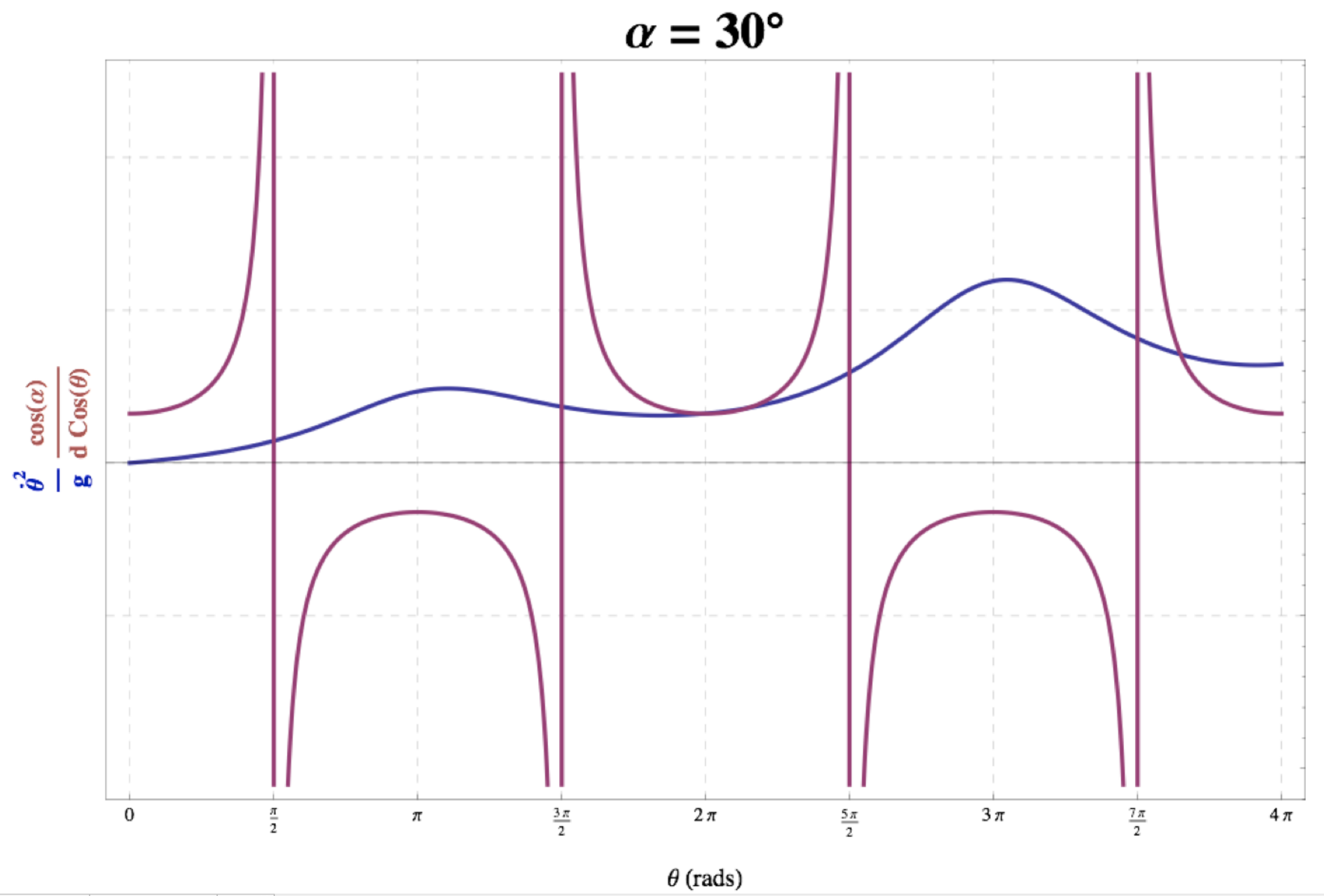}
\caption{Plot of $\dot{\theta}^2/g$ vs $\cos\alpha/d\cos\theta$, for $\alpha=30^\circ$¼. The first intersection is at $\theta =  11.46$ rads, so the cylinder jumps at $114.6$ cm.}
\label{f3}
\end{figure}

\begin{figure}[h!]
\centering
\includegraphics[width=13 cm]{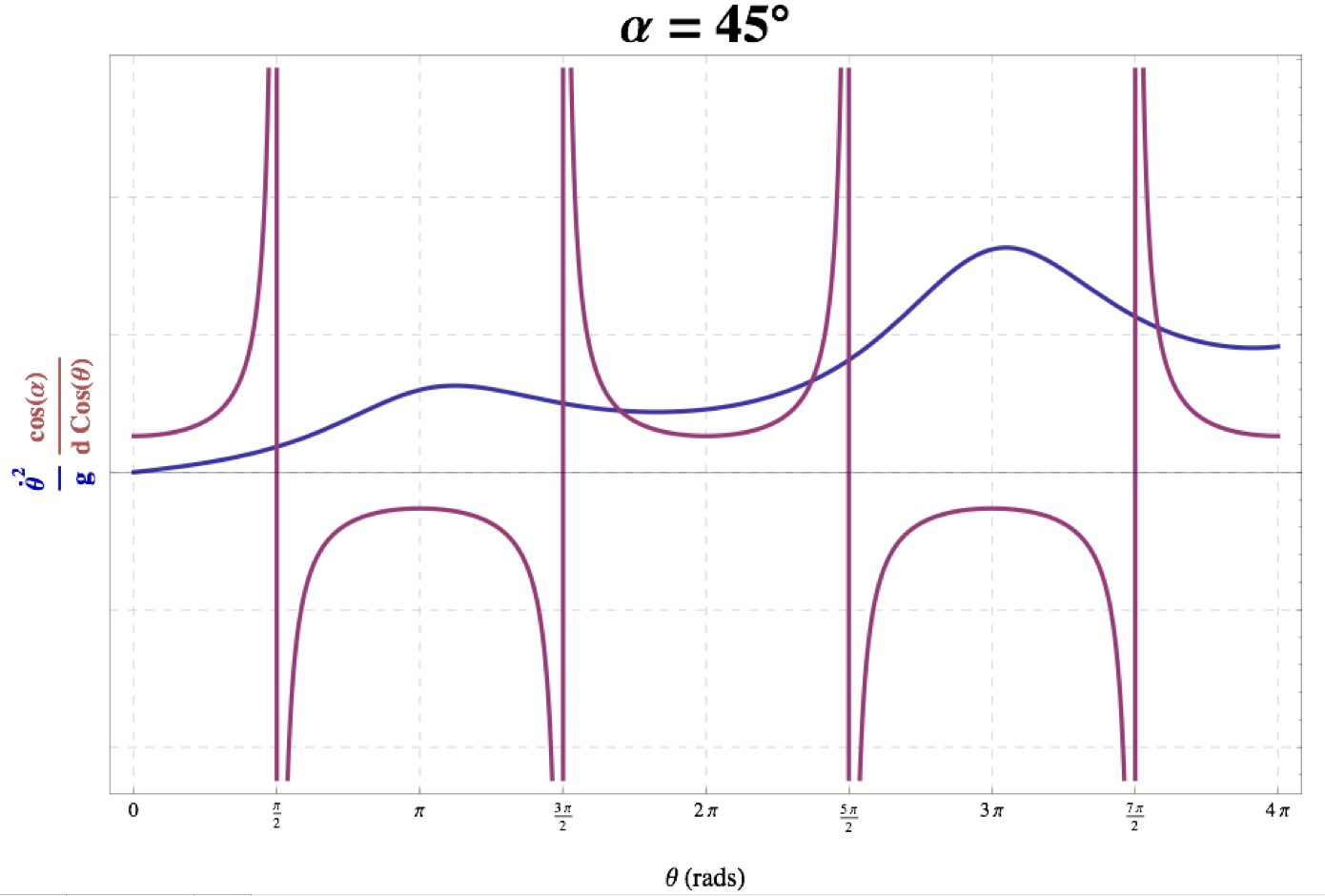}
\caption{Plot of $\dot{\theta}^2/g$ vs $\cos\alpha/d\cos\theta$, for $\alpha=45^\circ$¼. The first intersection is at $\theta =  5.34$ rads, so the cylinder jumps at $53.4$ cm.}
\label{f4}
\end{figure}

A simple way to arrive to a solution of this transcendental equation is plotting together both sides and looking for the first intersection of the resulting curves. The results for $15^\circ$, $30^\circ$ and $45^\circ$ are shown in figures \ref{f2}, \ref{f3} and \ref{f4}.

It is interesting to note the oscillation of the kinetic energy (proportional to the left side of Eq. \ref{E9}) due to the centrifugal force effect produced at different positions of the center of mass of the cylinder. 

\section{Experimental results}

To compare our theoretical results with experimental data, we use a Styrofoam cylinder of radius $r = 10.0 \pm 0.05$ cm, high $h = 5.55 \pm 0.05$ cm and mass $m_1 = 34.2 \pm 0.05$ g, to which a $9.50 \pm 0.01$ mm diameter and $5.10 \pm 0.001$ cm long brass road of mass $m_2 = 32.1 \pm 0.05$ g was imbibed parallel to the disc axis at a distance $y = 5.50 \pm 0.05$ cm from it, so the center of mass lies at a distance $d = 2.65 $ cm from the center of the cylinder. Then the disc rolls on a $3.20$ m long wooden ramp inclined at $15^\circ$, $30^\circ$ and $45^\circ$ with respect to the horizontal. To determine the jumping site, the motion was recorded with a high-speed video camera (Casio EX ZR100) at 240 and 400 frames per second. 
	
The main sources of errors in our experiment are the initial position of the cylinder, which was taken in such a way that its center of mass lied in the normal to the incline, and the place where the cylinder actually jumps. The initial position of the cylinder was done manually and we estimate an error of about $3^\circ$. The jumping site is difficult to determine in the video, especially in $45^\circ$ case, because the jump takes place before disk has completed a whole turn and its speed is relatively small, so the jump is minute. We estimate a maximum error of about $5.0$ cm. in readings of the jumping site. Taking this into account, no error analysis was attempted; instead we report our experimental results in the jumping site with a maximum error of $\pm 5.0$ cm. 

\begin{figure}[h!]
\centering
\includegraphics[width=15 cm]{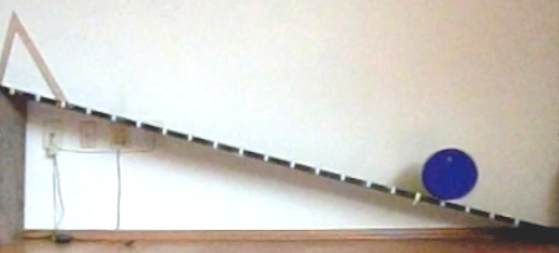}
\caption{Photogram at the jumping site for an incline at $\alpha = 15^\circ$.}
\label{f5}
\end{figure}

\begin{figure}[h!]
\centering
\includegraphics[width=14 cm]{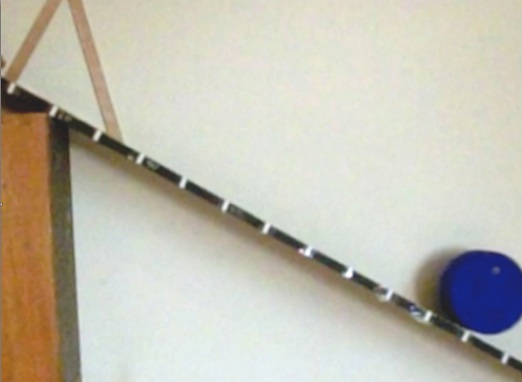}
\caption{Photogram at the jumping site for an incline at $\alpha = 30^\circ$.}
\label{f6}
\end{figure}

\begin{figure}[h!]
\centering
\includegraphics[width=16 cm]{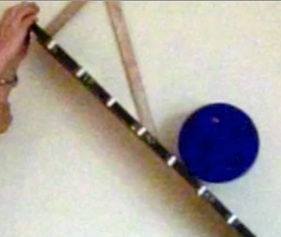}
\caption{Photogram at the jumping site for an incline at $\alpha = 45^\circ$.}
\label{f7}
\end{figure}

Figures \ref{f5}, \ref{f6} and \ref{f7} are photograms, taken from the videos, of the jumping sites and in table 1 we compare the theoretical solution with the experimental results.

\begin{table}[h!]
\begin{center}
\begin{tabular}{c||c|c|c}
 Incline angle $\alpha$ (degrees) & $15$ & $30$ & $45$ \\
\hline
\hline
 Theoretical jumping angle $\theta_\tau$ (rad) & $18.01$ & $11.46$ & $5.34$ \\
\hline
 Theoretical jumping position (cm) & $180.1$ & $114.6$ & $53.4$ \\
\hline
 Experimental jumping position (cm) & $185.0\pm5.0$ & $115.0\pm5.0$ & $55.0\pm5.0$ \\
\end{tabular}
\end{center}
\caption{Comparison of the experimental results with the theoretical predictions.}
\label{t1}
\end{table}

Of special interest (to be seen in the videos) are: the $15^\circ$ case, were the speeding and the slowing down of the cylinder are clearly seen, and the spectacular second and third jumps in the $45^\circ$ case. The videos can be seen in the following YouTube web address: 

\noindent $<$http://www.youtube.com/watch?v=ITQYHU2ekMM$>$ .

\section{Conclusions}

A first integral of the equation of motion of a cylinder whose center of mass is not at its geometrical center, and rolls on an incline, is obtained. From this solution the site where the cylinder should ÒjumpÓ is determine. An experimental setup, that resembles the assumptions made to obtain the theoretical solution, was furnished. The experimental results agree well with the theory. We believe that this is the first time that this problem is theoretically and experimentally addressed. 

\begin{acknowledgments}
This work was partially supported by DGAPA-UNAM IN115612, M\'exico.
\end{acknowledgments}

\end{document}